\documentclass[final,1p,times,twocolumn,authoryear]{elsarticle}

\usepackage{amssymb}

\usepackage{lineno}

\usepackage{marginnote}
\usepackage{color}

\journal{Tectonophysics}

\begin{document}

\begin{frontmatter}



\title{Aftershocks of the\\2012 Off-Coast of Sumatra Earthquake Sequence}


\author[a]{Chengping Chai\corref{cor1}}
\author[a]{Charles J. Ammon}
\author[b]{K. Michael Cleveland}
\address[a]{Department of Geosciences, The Pennsylvania State University, University Park, PA 16802, U.S.A.}
\address[b]{EES-17:Geophysics, Los Alamos National Laboratory, Los Alamos, New Mexico 87545, U.S.A.}
\cortext[cor1]{Corresponding author (chaic@ornl.gov), now at Oak Ridge National Laboratory, Oak Ridge, TN 37830, U.S.A.}

\begin{abstract}
Aftershocks of the 2012 Off-Coast of Sumatra Earthquake Sequence exhibit a complex and diffuse spatial distribution. The first-order complexity in aftershock distribution is clear and well beyond the influence of typical earthquake location uncertainty. The sequence included rupture of multiple fault segments, spatially separated. We use surface-wave based relative centroid locations to examine whether, at the small scale, the distribution of the aftershocks was influenced by location errors. Surface-wave based relative location has delineated precise oceanic transform fault earthquake locations in multiple regions. However, the relocated aftershocks off the coast of Sumatra seldom align along simple linear trends that are compatible with the corresponding fault strikes as estimated for the GCMT catalog. The relocation of roughly 60 moderate-earthquake epicentroids suggests that the faulting involved in the 2012 earthquake aftershock sequence included strain release along many short fault segments. Statistical analysis and temporal variations of aftershocks show a typical decay of the aftershocks but a relatively low number of aftershocks, as is common for intraplate oceanic earthquakes. Coulomb stress calculations indicate that most of the moderate-magnitude aftershocks are compatible with stress changes predicted by the large-event slip models.  The patterns in the aftershocks suggest that the formation of the boundary and eventual localization of deformation between the Indian and Australian plate is a complicated process. 
\end{abstract}


\end{frontmatter}

\section{Introduction}
Across most of the Earth, plate boundaries are well defined geologically and well delineated by seismic activity. The boundary between the Australian and Indian Plates is one of a few exceptions. Seismicity across the Indian Ocean Basin is characterized by diffuse activity that does not define a clear boundary \citep[e.g.][]{Wiens:1985fw, Delescluse:2007hj}. Some patterns are clear, for example a change in deformation style occurs from west to east across the Ninety-East (90E) Ridge - compressive deformation predominates to the west and strike-slip deformation dominates in the Wharton Basin to the east \citep{Delescluse:2007hj}. Lack of a clearly defined boundary does not preclude the occurrence of very large earthquakes. The 2012 Off the Coast of Sumatra sequence included two great events  \citep[e.g.][]{Yue:2012ks}. These large earthquakes and their aftershocks occurred in the northernmost Wharton Basin, which experiences NNW oriented compressive stresses as a result of slab pull associated with subduction of the Indian Ocean lithosphere along the Sunda Arc, the continental collision between the India and Eurasia plate to the north, and ridge push associated with Southeast Indian Ridge \citep{Delescluse:2007hj}. The intraplate region has hosted numerous large earthquakes \citep[e.g.][]{Antolik:2006fh,  Abercrombie:2003fd, Aderhold:2016dd, Lay:2016ba}, none more spectacular than the 11~April 2012 sequence that began with an $M_W$~7.3 earthquake in January and culminated with two great earthquakes, an $M_W$~8.6 and an $M_W$~8.2, separated spatially by about 200 km and temporally by about two hours \citep[e.g.][]{Duputel:2012ey, Zhang:2012gx, Meng:2012gk, Wei:2013fv, Ishii:2013bp, Wang:2012jk}. The events locate westward of the recent large-magnitude megathrust activity that began with the 2004 Great Sumatra-Andaman Island ($M_W$~9.2) earthquake \citep[e.g.][]{Ammon:2005hz, Lay:2005ey}.

Early analyses of the 2012 aftershock patterns and rupture identified substantial complexity and indicated that significant slip must have occurred along numerous faults \citep[e.g.][]{Yue:2012ks}. The sequence is special in many ways and occurred several hundred kilometers seaward of the active plate boundaries. The 11~April, 2012 $M_W$~8.6 \citep[Global Centroid Moment Tensor, GCMT, ][]{Ekstrom:2012by} earthquake is both the largest strike-slip faulting event and the largest intraplate earthquake that has been seismologically recorded \citep{Meng:2012gk, McGuire:2012if}. Two hours later, an $M_W$~8.2 strike-slip earthquake struck the same general intraplate region. Multiple fault segments \citep[e.g.][]{Yue:2012ks} participated in the compound large-event sequence. Previous studies debated the basic parameters of the $M_W$~8.6 mainshock. A rupture speed of 2 km/s was inferred by \citet{Yue:2012ks}, although a later study claimed supershear-rupture of ~5 km/s \citep{Wang:2012jk}. Notable differences exist in the fault systems inferred in early investigations \citep{Meng:2012gk, Wang:2012jk, Yue:2012ks}, but all agree that the rupture was complicated and included strain release on numerous structures. 

Now, just over five years following the main strain release, information from aftershocks can be exploited to explore the intraplate structures activated during these impressive earthquakes. In the years since the mainshocks, aftershock activity has continued and the initial, complicated seismicity patterns have become slightly, but not completely well defined. The United States Geological Survey (USGS) catalog includes epicenter locations and depths for several hundred of events. Although of limited spatial extent, high-resolution bathymetric investigations provide some interesting clues to the involvement of en echelon fracture-zone reactivation and of conjugate fracture-zone and shear-zone participation in deformation across the region \citep[e.g.][]{Carton:2014gq,Qin:2015ek,Singh:2017gi}. In this work, we explore the patterns in the USGS earthquake locations and the GCMT faulting geometries, and then use a surface-wave-based precise, relative epicentroid location technique \citep{Cleveland:2013dm, Cleveland:2015gh} to better quantify spatial relationships of moderate-magnitude events within several of the key subclusters that comprise the overall complex aftershock pattern of this important earthquake sequence.
 
\section{Earthquake Catalog Aftershock Patterns}
We examined the seismicity in the USGS ComCat and International Seismological Centre (ISC) catalogs, and the faulting geometry in the GCMT Catalog to investigate the aftershock sequence of these great oceanic, intraplate earthquakes. We begin with a descriptive review and then integrate our relocation results with catalog information later in the discussion.

\begin{figure}[!ht]
\centering
\includegraphics[width=\textwidth]{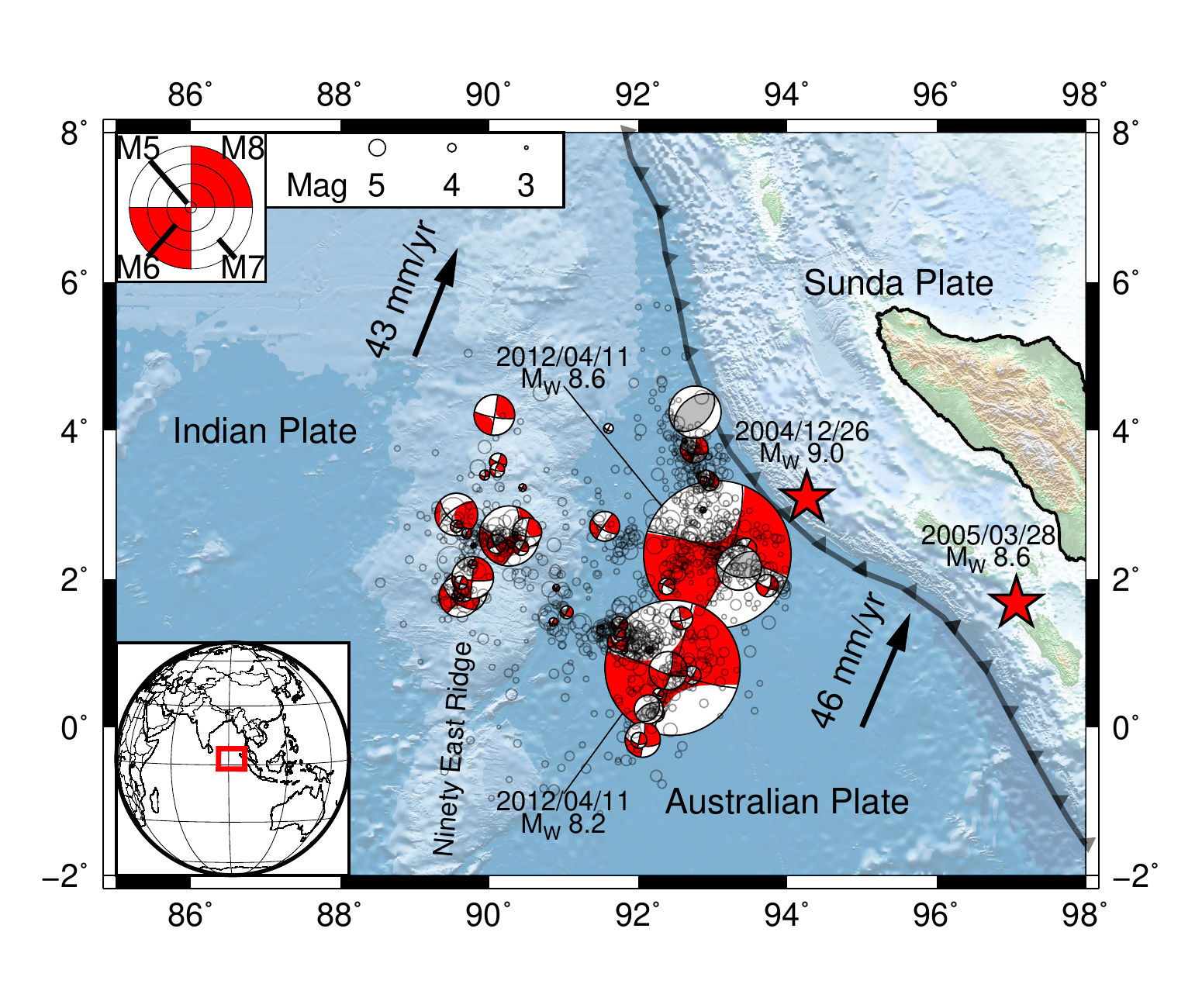}
\caption{Regional seismicity of the off-Sumatra region between January 2012 and February 2016.  Focal mechanism from GCMT catalog are shown with red (strike-slip and one normal faulting) and gray (reverse faulting) beach balls. Magnitude 3 and larger earthquakes from ISC catalog \citep{ISCcitation2010} are shown as open circles. Two stars indicate the GCMT centroids of the 2004 $M_W$ 9.0 and the 2005 $M_W$ 8.6 earthquakes. The lower insert shows the location of study area on a global map as a red box. Upper inserts indicate the magnitude scale used for symbol sizes. Black arrows show plate motions from MORVEL \citep{DeMets:2010dn}. Plate boundaries (black line with triangles) are based on \citet{Bird:2003ip}. }
\label{fig_gcmtMechanismBaseMap}
\end{figure}

The 2012 aftershock pattern reflects the complexity of the initial faulting, and lacks any clear indication of long simple structures originating near the epicenters of either of the two great earthquakes (Fig.~\ref{fig_gcmtMechanismBaseMap}). Instead, the aftershocks spread within a two-dimensional pattern stretching from near the Sumatra Trench to the Ninety-East Ridge and between the latitudes of roughly 1.5N and 4.5N. The region is covered by several kilometers of Bengal-Fan sediments that obscure all but the largest bathymetric features that could provide more clues to the structures participating in the deformation. The 2012 aftershock zone has linear dimensions on the order of 500 km and encompasses roughly 200,000~$km^2$. Many of the aftershocks occurred within the first week, immediately showing the complexity of the strain release of the large events and the complex strain distribution within the Indian-Ocean lithosphere. Early strain release (first day) covered an area roughly 130,000~$km^2$. The evolution of the pattern with time has clarified the overall geometry of a number of the subclusters. But events within each cluster remain diffuse in some regions, only partly a result of uncertainties in event location. 

\begin{figure}[!ht]
\centering
\includegraphics[width=\textwidth]{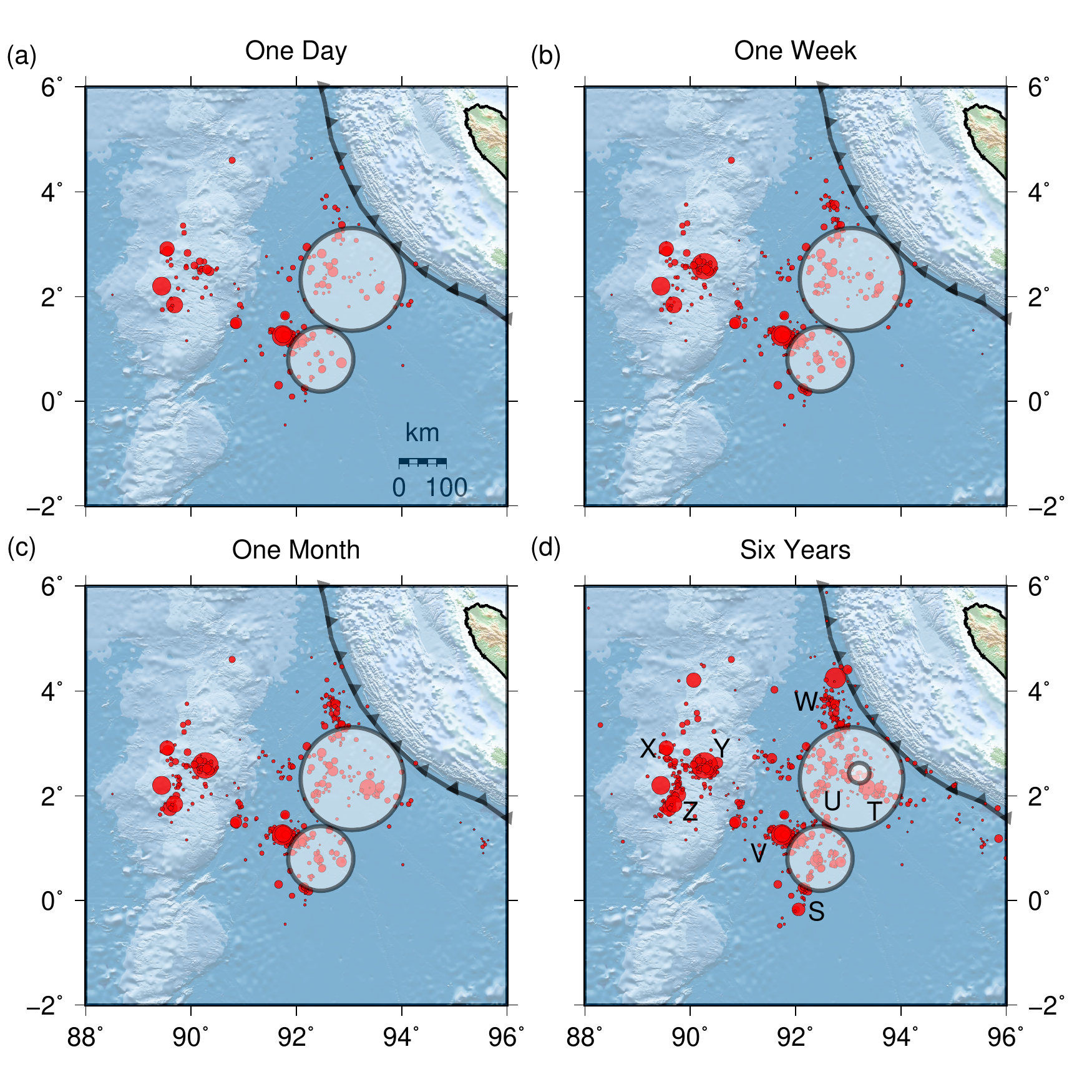}
\caption{Maps of USGS seismicity showing intraplate events in the vicinity of the 2012 earthquake sequence. Events with magnitudes less than 7 are identified by circles scaled logarithmically to four times the expected rupture area for an earthquake with the corresponding magnitude (assuming a simple earthquake model). Epicenters for larger events are indicated by white circles scaled to roughly the rupture area expected for their magnitudes. The panel in the upper left shows the events occurring within one week of the mainshock ($M_W$ 8.6); the upper right shows events occurring within one month of the mainshock, events in the lower left occurred within one year of the mainshock; the lower right panel shows seismicity from the start of January, 2012 through April, 2018. }
\label{fig_usgsAShockMaps}
\end{figure}

USGS epicentral locations for roughly 730 events within the region occurring since 01~January, 2012 are shown in Fig.~\ref{fig_usgsAShockMaps}. We extracted hypocenters from the catalog within a polygon defined by the epicentral patterns extending west from the accretionary prism to and including the 90E Ridge within the rectangular region bounded by corners at (-7N, 84E) to (7N, 96E). The figure shows aftershock patterns as they evolved from one week to one month to one year following the event. The fourth panel shows the 2012 foreshocks and all events in the sequence into April 2018 (roughly six years).  Most of the area that would eventually be involved in the strain release was active within the first week, and the difference between one month and one year is minimal. The USGS event depth estimates are dominated by 511 events with fixed depths at 10~km, but of the 212 remaining all are shallower than 40 km, and most are shallower than 35~km. However the great majority of earthquakes without fixed depths are located below the oceanic crust. 

Over the last six years some lineations have become more apparent within the aftershocks, including (S, Fig.~\ref{fig_usgsAShockMaps}d) one stretching about 140-150~km with an azimuth of roughly 25N, which is not that different from the GCMT point source model strike of 17N for the $M_W$~8.2 event. As is commonly the case in remote regions, the trend is not precisely defined by the epicenters and considerable spread in the epicenters leaves open the possibility that the trend is formed by numerous faults in an en echelon formation. Even more diffuse is a roughly 200 km pattern (T) including the $M_W$~8.6 epicenter oriented in a direction of roughly 314N, which corresponds roughly to the plane striking 289N in the GCMT point source model. A 90~km long lineation (U) extends in a direction of 30N and intersects the NW-SE mainshock trend about 50~km from the mainshock. A spatially distributed cluster of activity (V) about 190~km southwest of the mainshock epicenter strikes in a direction of 300N for about 100~km but is also about 40 km wide orthogonal to that trend. A cluster north of the mainshock (W) strikes roughly 350N and extends for roughly 130 km. Far to the west, several clusters appear to have ruptured features within the 90E ridge, an area of hot-spot modified oceanic lithosphere that corresponds to one of the most dramatic bathymetric features on Earth. Sub-clusters of that region include a roughly 100 km long segment (X) striking 300N that intersects an 85-km long segment striking 30N (Y), and a region that can only be described as a 40 by 50 km zone of seismicity (Z). In addition to these relative large features, several smaller clusters are apparent.

Catalog earthquake locations in the surrounding region and the GCMT solutions for the two great earthquakes are included in Fig.~\ref{fig_gcmtMechanismBaseMap}. Seventy-six events from the aftershock region are listed in the GCMT catalog. Thirty-one events had fixed centroid depths (15 or 12 km), 45 had free depths. Of the 45 earthquakes with depth estimates, the depths range from about 12 to 55~km, and most events had depths shallower than 35 km, with the notable exceptions of the two great events which had point-source centroid depth estimates in the range of 45 to 55~km. No obvious spatial pattern is associated with the depths. Most of the events are strike-slip in nature and the compressional direction associated with the deformation is NNW and the tension direction is ENE. A handful of reverse faulting events are included in the aftershock sample and they generally are near the Sumatran subduction zone (they formally locate outboard of the trench) and show trench-perpendicular compression. The compressional direction from the GCMT faulting geometry estimates is consistent with sea-floor-bathymetry-constrained orientations of normal faults within the shear zones conjugate to the fracture zones.

Relocation of all aftershocks is an obvious long-term goal, but acknowledging that the best relocated events will be the largest events, which have the most observations and the best azimuthal coverage, we focus on relocating moderate and strong aftershocks along with the major foreshock and $M_W$ 8.2 great event. Specifically, we use surface-waves to relocate the moderate and large events in the sequence. Although we investigate more than fifty event locations, this number is inadequate to define broad patterns within the aftershock sequence. Still, the moderate-magnitude events provide clues to the internal structure of several of the interesting aftershock clusters apparent in the less precise USGS aftershock locations. We focus on sixty-one strike-slip earthquakes located within the aftershock clusters to investigate whether the diffuse nature of the overall seismicity pattern continues to smaller spatial scales within each of the clusters.

\section{Surface-Wave-Based Relative Earthquake Location}

Since we are going to use waveform similarity to measure relative travel time differences for signals from different events, we must ensure that the earthquakes have similar faulting geometries \citep{Cleveland:2013dm}. Closely located earthquakes show similar waveforms at most stations. Example waveforms from two earthquakes are shown in Fig. S1 of the electronic supplement. We searched the Global CMT catalog for earthquakes in the focus region. Most of the off-Sumatra aftershocks are strike-slip events, one is a normal faulting event and four are reverse faulting events. We focus our effort on a sixty-one strike-slip faulting earthquakes. Relative surface wave arrival times can be complicated by changes in strike and dip of the strike slip events, but we examined all the azimuthal time-shift patterns to ensure the data used for the relocations are consistent.  For the most part, time difference from nearby event pairs show a consistent sinusoidal variation, which is associated with the relative position of the event epicentroid \citep{Cleveland:2013dm}. We have included two examples (Fig. S2 and Fig. S3) in the supplementary document. The sinusoidal variation can be clearly seen in the figures.

\begin{figure}[!ht]
\centering
\includegraphics[width=\textwidth]{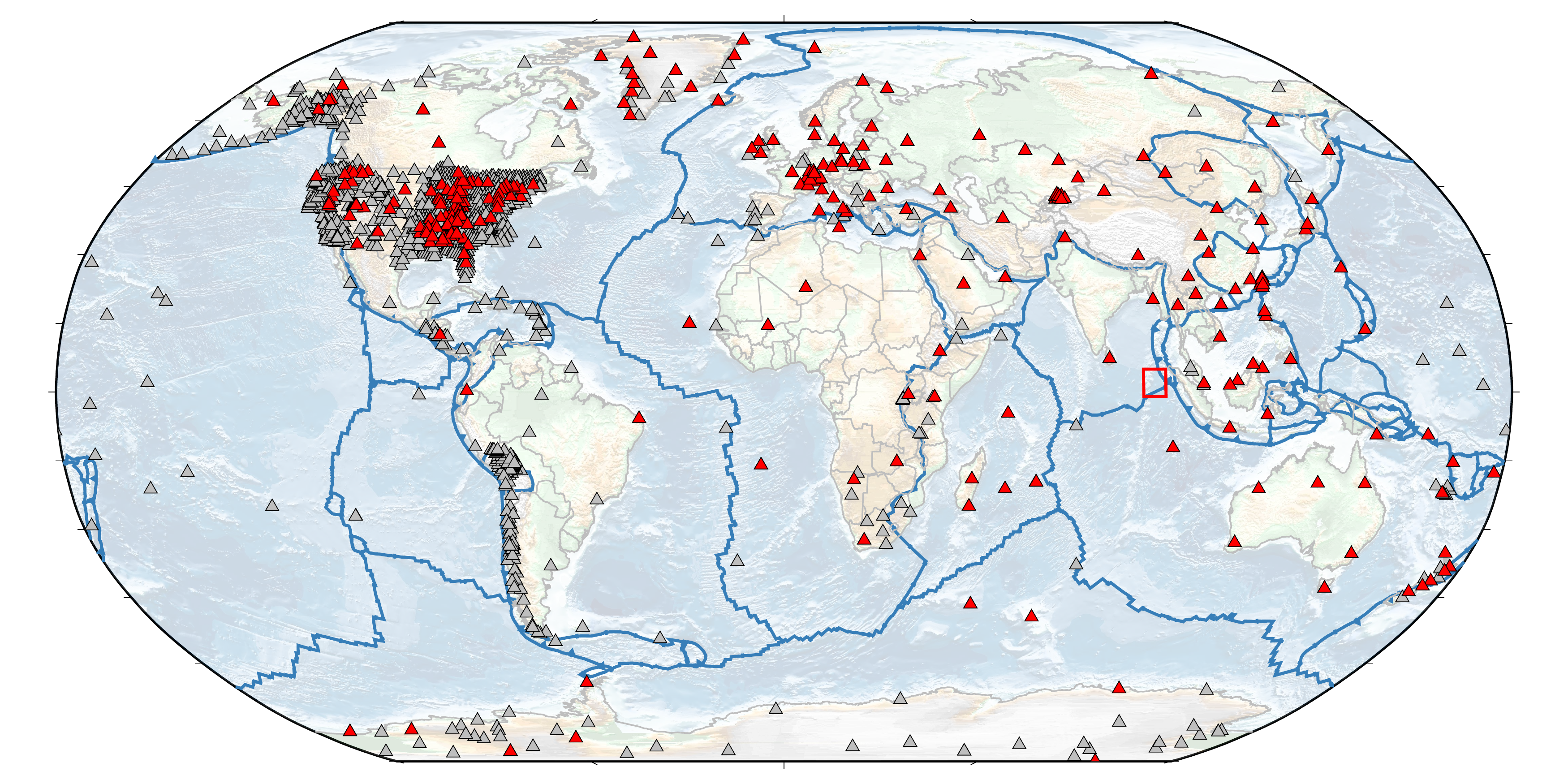}
\caption{A Map showing the seismic stations tried (gray triangles) and used (red triangles). The red box indicates where the earthquakes of this study are located. Plate boundaries (blue lines) are based on \citet{Bird:2003ip}. }
\label{fig_station_map}
\end{figure}

Our seismogram analysis procedure is similar to \citet{Cleveland:2013dm}. We correlated short-arc Rayleigh waves and Love waves seismograms to measure the relative time shifts between similar waveforms. We used intermediate period (30-80~s period) signals recorded at stations operating at the time of the event and downloaded from the Incorporated Research Institution for Seismology (IRIS) Data Management Center (DMC).  Fig.~\ref{fig_station_map} shows the seismic stations we used for this study. The instrument-response was removed from the seismograms through a frequency-domain deconvolution. The bandwidth is chosen for several reasons: first, the group slowness in this band is relatively constant for oceanic lithosphere, which simplifies the location analysis; second, the noise in this band is relatively low; third, the period range is relatively insensitive to modest changes in source depth (particularly for strike-slip sources); fourth, the long wavelengths allow linking across relatively long distances.  All waveforms were visually inspected and assigned a quality grade based on the signal-to-noise ratio and character of the surface waves. The selected waveforms were graded from A (best) to F (worst) by visual inspection of the signal quality. Only waveforms with quality better than C are used for further analysis. Examples  of waveforms with different quality grades are shown in Fig. S4. 

The surface waves were isolated from the seismogram using a group slowness window of (0.2 to 0.4~s/km) and signals from nearby (within 150 km radius) events recorded at the same station were cross correlated to estimate the relative time shifts and to form an interconnecting network of event pairs. Based on previous  studies \citep{Cleveland:2013dm, Cleveland:2015gh}, only waveforms having normalized cross-correlation values larger than 0.9 from two similar events are used in the relocation. The relative time-shifts from the higher quality waveforms are used in a linearized inversion of double-difference time shifts to constrain relative earthquake epicentroid locations. Surface-wave cross correlation time shifts are relatively insensitive to modest depth differences. Since we are using signals with wavelengths of several hundred kilometers, our inversion is in terms of the event spatial and temporal centroids and we use the term epicentroid to represent the location on Earth's surface above the rupture's spatial centroid. For simplicity, we refer to the time shifts as origin-time shifts, since most events are small and the centroid and origin times are close. Double differences are calculated for all linked events and an iterative, nonlinear, inversion for the change in latitude, longitude, and origin time is solved using a truncated SVD. Double-difference partial derivatives are computed assuming a uniform slowness (spatially and for the entire bandwidth). We usually select the slowness value by using a grid search and choosing the value that produces the smallest changes in location relative to the original (USGS) locations. This often leads to only slight changes in the total centroid of all the epicenters. The inversion includes no direct absolute location constraints, but in practice the initial locations (USGS epicenters) provide some a priori information. For inversion result assessment, individual event-pair patterns were visually inspected to check the quality of the data, which should exhibit a sinusoidal pattern for events with similar mechanisms and depths. 

\begin{figure}[!ht]
\centering
\includegraphics[width=\textwidth]{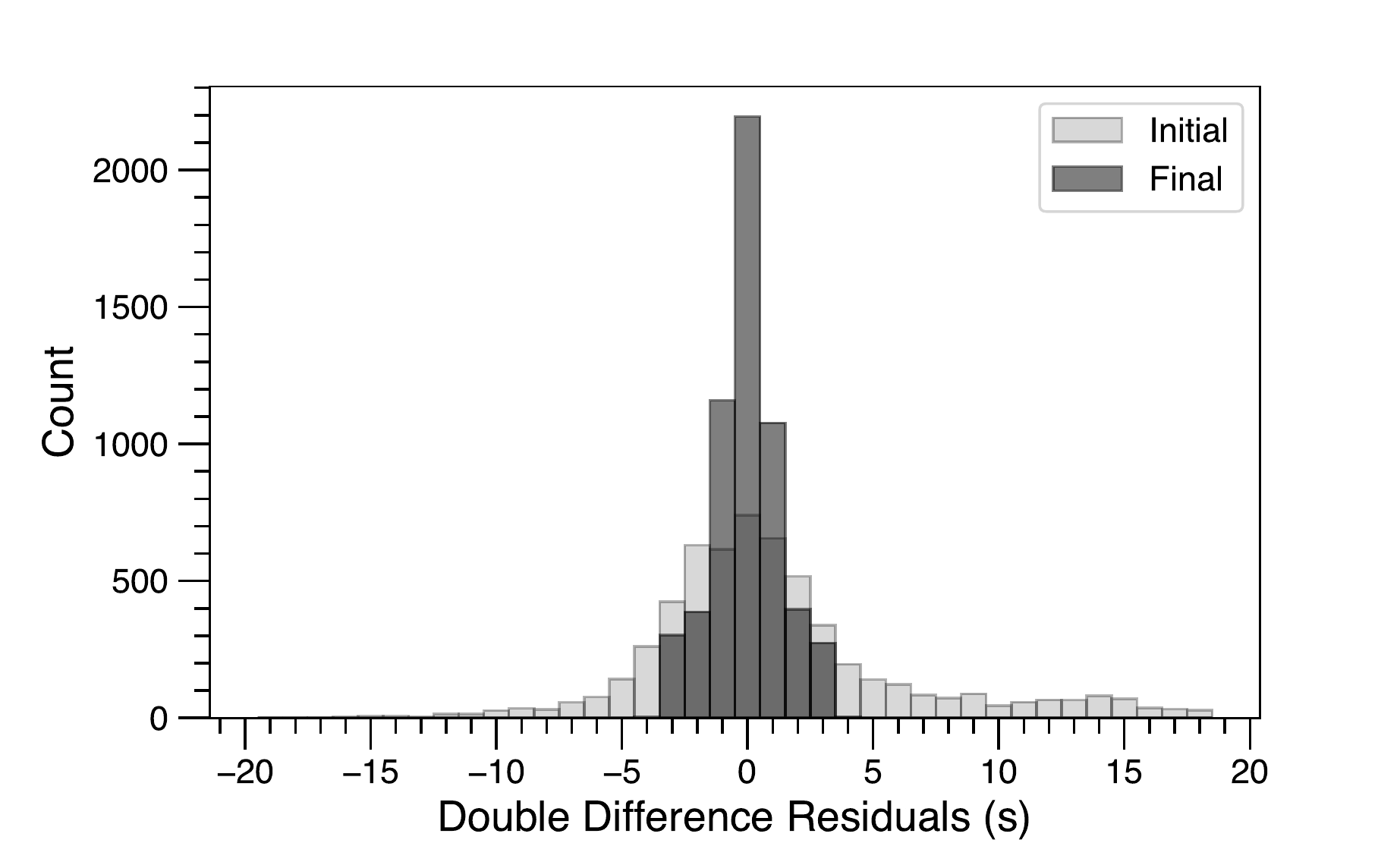}
\caption{Initial and final misfits for the inversion of observations from 61 strike-slip earthquakes. Initial misfit is computed based on USGS epicenter locations and origin times.}
\label{fig_residualHistogram}
\end{figure}

The inversion for relative earthquake locations is the same as that used in \citet{Cleveland:2015gh}. A spherical-earth version of the double-difference relocation approach \citep{Waldhauser:2000je} is applied to measurements from short-arc surface waves. A linearized inversion is constructed relating the observed and predicted surface-wave arrival time differences to changes in earthquake epicentroid position and origin time.  A constant slowness is assumed since the first-order seismic structure variations between events are negligible for the 30-80 s period range in which time shifts are measured. We used slownesses of 0.257~s/km for Rayleigh wave and 0.223~s/km for Love wave measurements based on values from a global dispersion model \citep{Ekstroem:2011gb}. We tested the results using a range of reasonable slowness values to verify that the general patterns in the relocations are only mildly sensitive to the specific assumed slowness. Larger slowness values produce larger location shifts from the initial locations, lower slowness values lead to smaller location shifts. Similar to \citet{Cleveland:2015gh}, we allow earthquakes within 150 km to be linked as long as at least 12 common stations are available. A linking criteria based on the azimuth distributions of stations was also applied to assure linked events have a sufficient azimuth coverage. The azimuth coverage is reasonable for the linked events. The station coverage for two of the linked events is shown in Fig. S5.  After three iterations, the misfit to observations improved significantly as shown in Fig.~\ref{fig_residualHistogram}. The RMS misfit for the final locations is about 1.3~s, reduced from the initial misfit of about 5.2~s. We visually examined data fits for all the linked event pairs. The azimuthal coverage is not optimal for some smaller events due to sparse station coverage to the south. However, a recognizable cosine pattern exists for most events and we ignore poorly constrained events in our subsequent discussion. 

To verify the relocation results, we performed several synthetic tests similar to \citet{Cleveland:2013dm} and \citet{Wang2018}. Assuming the station coverage is the same as that in Fig. S5, surface-wave seismograms for five hypothetical earthquakes were simulated with the Computer Programs in Seismology package \citep{herrmann2013} using the AK135 velocity model \citep{kennett1995}. We assume these five Mw 5.5 earthquakes were located on a linear fault striking in the N-S direction. The seismograms were computed with the true locations of these earthquakes. Incorrect locations were then assigned to these seismic events. The double-difference relocation algorithm was used to improve the earthquake locations from the incorrect ones. For the first synthetic test, we assume all the earthquakes have the same strike-slip focal mechanism and focal depth (5 km). The relocation results and a time-shift plot are shown in Fig. S6 and S7, respectively. Though the absolute locations of these earthquakes are shifted together to the southeast in respect to the true locations, the relative locations of the earthquakes are well recovered. If we shift the centroid of the inverted locations to the centroid of the true locations, the difference between the inverted locations and the true locations are around 1 km on average. For the second test, we assume a different focal depth (5, 10, 15, 20, and 25 km) for each of the five earthquakes. The relocation results are almost the same as those for the first test. For the third, forth and fifth synthetic test, the dip, rake, and strike of the five earthquakes was perturbed based on distributions measured from real data in the region. We found that variations in strike does not change the relocation results. Perturbations on dip and rake leads to larger errors in the location. However, the inverted locations are much better than the initial locations. Ignoring the centroid shift, the location errors are less than 2 km for all of these five synthetic tests. The location errors are around 4 km if we account for the centroid shift, which agrees with results by \citet{Wang2018}.

\section{Results}

\begin{figure}[!ht]
\centering
\includegraphics[width=\textwidth]{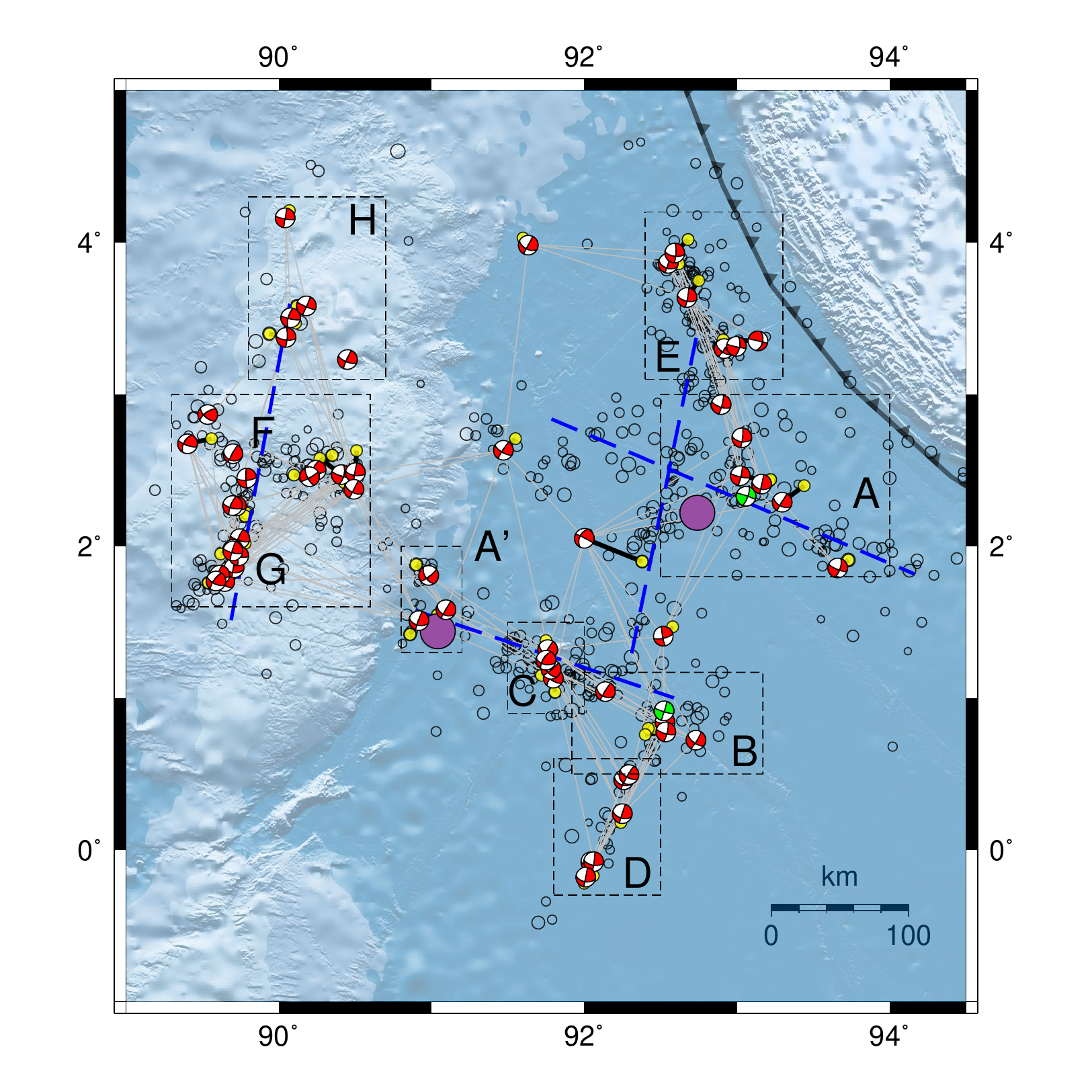}
\caption{Relocated epicentroids (red beach balls) from this study compared to original USGS locations (yellow dots). Black lines indicate location shifts between relocated epicentriods and USGS locations. The green beachballs are the two great earthquakes occurred on April 11, 2012. Red beachballs represent other events used in the relocation. Gray lines show the event links. Open circles show earthquake locations from USGS catalog. The boxes (dashed line bounded) indicate locations for 8 sub clusters. The large solid dots show the two-point locations associated with the $M_W$ 8.6 event proposed by \citet{Duputel:2012ey}. The thick dashed lines indicate fault segments used by \citet{Hill:2015iw}. }
\label{fig_clusterMap}
\end{figure}

\begin{figure}[!ht]
\centering
\includegraphics[width=\textwidth]{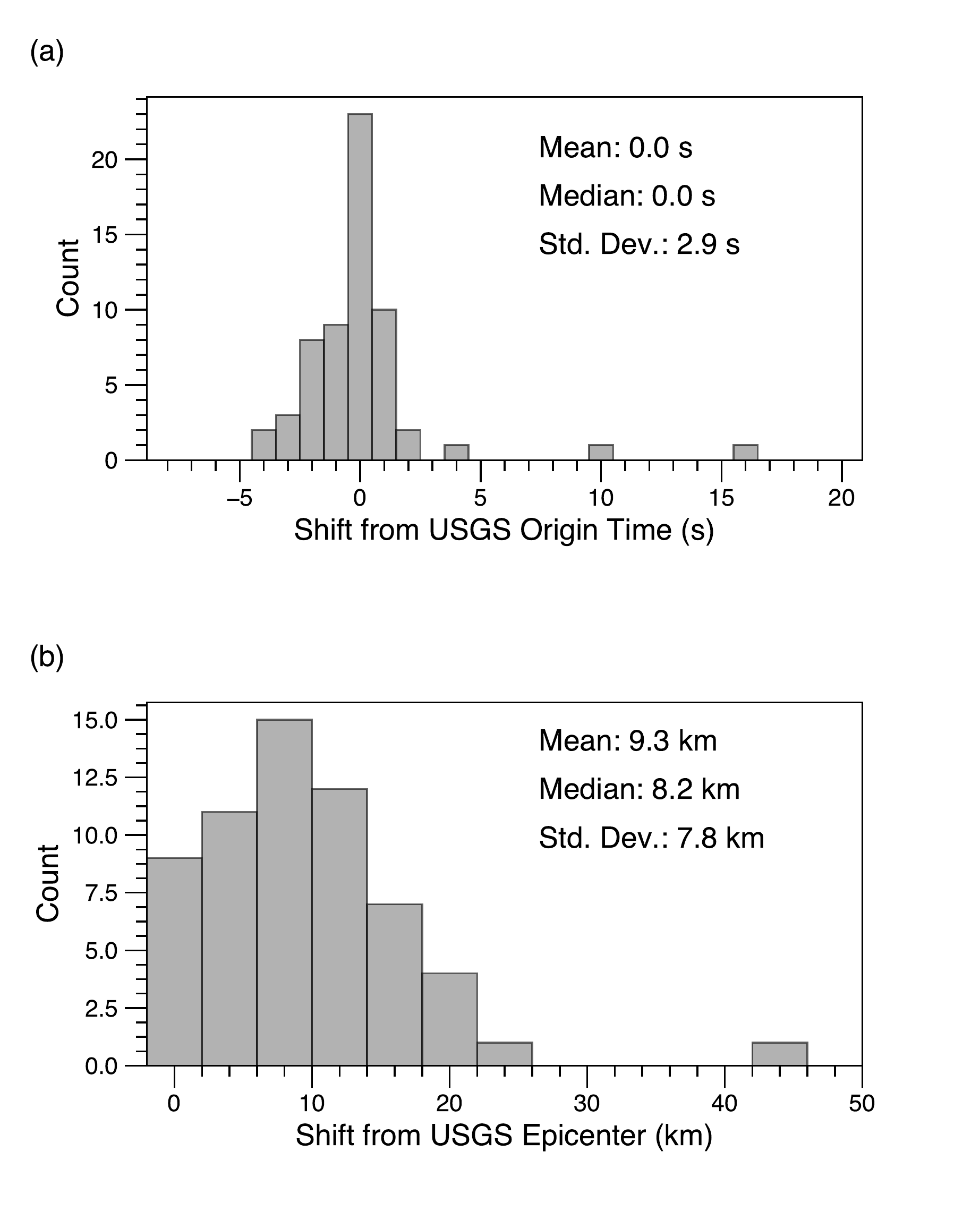}
\caption{Histograms show comparison of the epicentroid locations with the original USGS locations.}
\label{fig_locationShiftHistogram}
\end{figure}

Fig.~\ref{fig_clusterMap} is a map of the relocated epicentroids (center of focal mechanisms) and the original USGS locations (yellow circles). At this scale, the difference in many locations is modest but as described above, the new locations fit the observations significantly better than the initial locations. Since the location changes are too small to show for the entire study area on one map, we separate the relocated events into clusters and discuss each individually.  We summarize the overall location shift from the USGS catalog in Fig.~\ref{fig_locationShiftHistogram}. Origin times after relocation did not change dramatically except for a few events and the average origin time shift from the USGS catalog times is zero.  With one exception, the relocated epicentroids shifted less than 26 km. The 29~April, 2012 event ($M_W$~5) was shifted 46 km. Even for that event, located near the center of the study area, the visual fit to the observed sinusoidal time shift pattern (with azimuth) after inversion is good. On average the epicentroid locations shifted about 10~km from the initial USGS positions (the median shift was 8~km). The standard deviation of the location shift is 8~km. The location changes are about two times smaller than similar investigations of earthquakes along oceanic transform faults \citep{Cleveland:2013dm, Cleveland:2015gh}. We computed formal least-squares uncertainties estimates \citep{Kintner:2018kd}  assuming uncorrelated and a uniform data variance of two seconds (double our final RMS misfit, to be conservative). Most estimated uncertainties are less than three kilometers and the less well-constrained epicentroids are roughly twice that in both latitude and longitude directions. While these formal uncertainties are useful for identifying the better constrained data, they are approximate since they exclude information on faulting geometry and slowness uncertainties. The estimated uncertainties are comparable to what has been found at oceanic transform fault boundaries, where bathymetry corroborates the results. Including insight gained from experiments with varying the slowness and focal parameters, we feel that less than five kilometers is a reasonable estimate of the relative epicentroid uncertainties.

To explore the relocations, we examine eight event clusters identified in Fig.~\ref{fig_clusterMap}. Maps for the individual clusters are included in the online supplement. Little of the aftershock region has detailed bathymetric information making simple bathymetric validation of the trends difficult. Only one region has detailed bathymetry, so we begin our review with Cluster B. 

\subsection{Cluster B}
\begin{figure}[!ht]
\centering
\includegraphics[width=\textwidth]{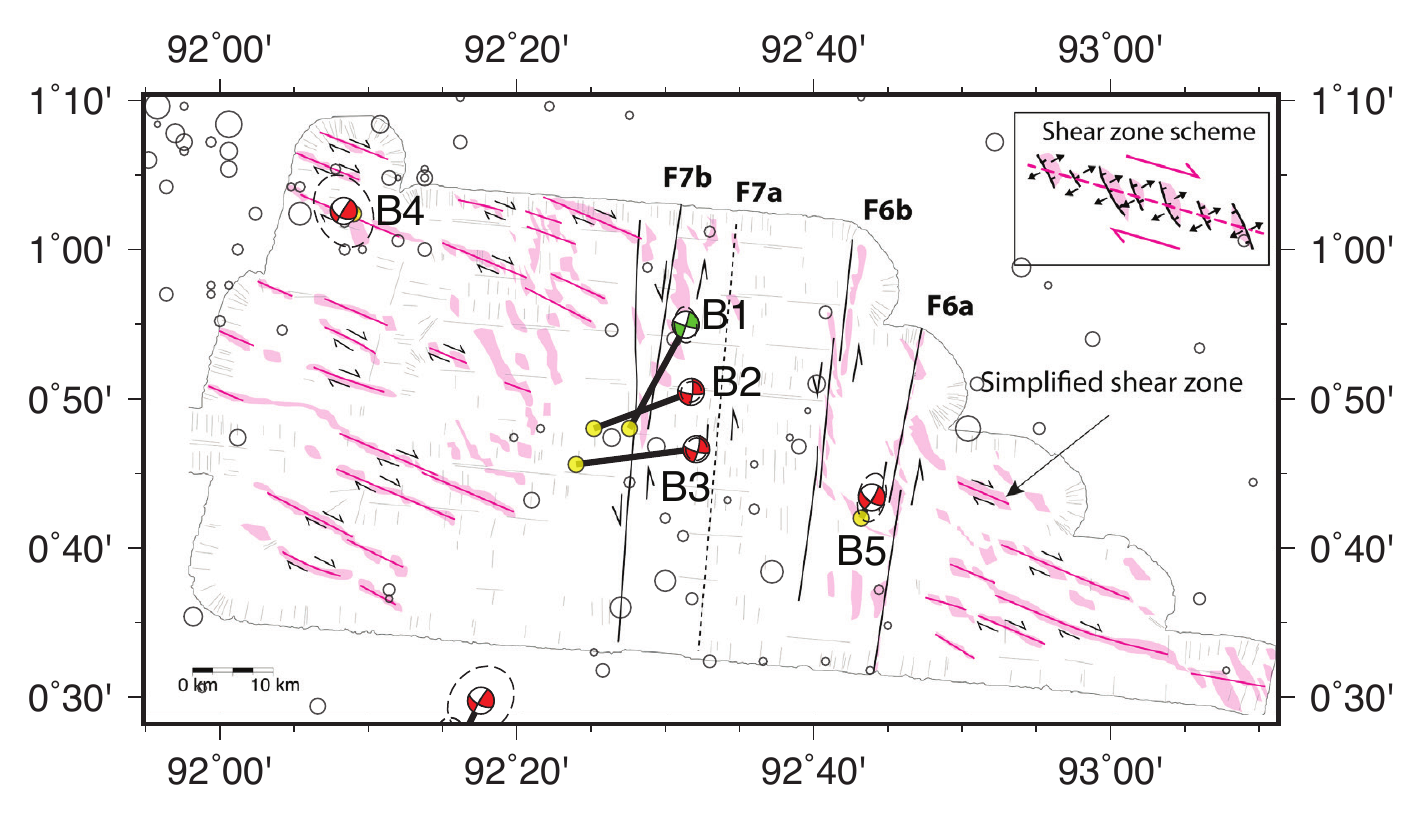}
\caption{A close view of the relocated epicentroids (red beachballs) around the 2012/04/11 $M_W$ 8.2 earthquake (the green beachball). The background is modified from \citet{Singh:2017gi}. The location of this map is shown as box B in Fig. \ref{fig_clusterMap}. Thick black lines indicate location shifts between relocated epicentriods and USGS locations (yellow dots). Thin black lines and the dash line indicate reactivated fracture zones. Red lines show shear zones. Ellipses indicate estimated uncertainties of earthquake locations. The inset shows a schematic shear zone scheme.}.
\label{fig_clusterB}
\end{figure}

Cluster B locates in the only region with high-resolution bathymetric information. We can use this information to test the consistency of locations similar to what was done by \citet{Cleveland:2013dm, Cleveland:2015gh} using oceanic transform related bathymetry. In Fig.~\ref{fig_clusterB}, we overlay background seismicity and relocated earthquakes onto a fracture- and shear-zone map that was derived from high-resolution bathymetry and a reflection survey \citep{Singh:2017gi}. Although we have limited absolute location information, event B1 ($M_W$ 8.2) locates in the vicinity of large fractures zones mapped and inferred along the sea floor.  Only three relocated events show any indication of nearby alignment - B1 (2012/04/11, $M_W$ 8.2), B2 (2012/04/16, $M_W$ 5.3) and B3 (2015/05/14, $M_W$ 5.7). The GCMT faulting geometry for each of the three events has a plane well aligned with the fracture zone strikes. However, the epicentroid trend does not align with either focal mechanism strike, nor the fracture-zones, and instead suggest en echelon fault ruptures roughly parallel to the overall fracture-zone trends. Teleseismic body-wave back-projection results for the $M_W$ 8.2 (B1) event suggest northward rupture along two nearby structures \citep[e.g.][]{Yue:2012ks}. A nearly north-trending rupture activating several en echelon faults can explain relocated aftershock locations, GCMT focal mechanism, bathymetry and reflection survey and back-projection observations. The depth of these three events (17-25 km, USGS NEIC) are well beneath oceanic crust though the uncertainties in depth estimate are large. Surface waves from the $M_W$~8.6 main shock obscured body wave observations from the $M_W$~8.2 event making waveform estimation of the depth challenging. Event B4 is offset east from the three-event cluster by about 20 km, which may indicate that as much as 20-30 kilometers of nearby fracture zones participated in the large event (the region between the roughly east-west shear zones in the bathymetric model). Event B5 is off to the west, and appears consistent with seismic failure along the conjugate shear zones identified by \citet{Singh:2017gi}. The unlabeled event to the south is discussed with the cluster D, which is south of this region. Changing the slowness moves the events  slightly, but never moves events B1-B4 out of the fracture zone region, nor B5 from the shear zone environment, and does not alter any of the relative location trends of the events. The biggest impact is on repositioning the cluster centroid by less than about 10-15 km.

\subsection{Cluster A}
Subcluster A (Fig. S8 in the electronic supplement) locates in the epicentral region of the $M_W~8.6$ mainshock. Surface waves from the large event do not correlate with the moderate-size event signals because of the substantial difference in frequency content for events of such different seismic moment. Thus we cannot provide precise locations of events relative to the $M_W~8.6$ mainshock (green focal mechanism). In fact, this region of substantial mainshock strain release has hosted only five moderate-magnitude strike-slip events. Five of the six events seem to align along an arc oriented NNW roughly in the direction of compression indicated by the GCMT faulting geometries and almost 45$^\circ$ off the strike of either plane in the GCMT focal mechanisms. Teleseismic body-wave backprojection and finite-fault models \cite[e.g.][]{Yue:2012ks} suggest failure of a large structure oriented closer to the GCMT strike directions than the sparsely aligned epicentroid locations in this cluster. We do not think these epicentroids identify a single significant structure. Our locations again suggest en echelon faulting, and show little evidence of aftershock activity on any large scale structure in the $M_W~8.6$ epicentral region. 

\subsection{Cluster C}
Short and long period analyses \citep{Duputel:2012ey} suggest that significant strain release also occurred in the region of Cluster~C and A' (Fig. S9 in the electronic supplement) during the $M_W~8.6$. Relocation brings the four moderate-size aftershocks within Cluster C into a near-linear alignment, but again the alignment is in a direction (NNW) that is at odds with the individual event faulting geometries and the average faulting geometry of the mainshock. The sparse number of moderate and larger magnitude events again suggests that the aftershocks are occurring along en echelon structures adjusting to mainshock slip, not illuminating a principal structure involved in the large event. 

\subsection{Cluster D}
Subcluster D (Fig. S10 in the electronic supplement) is presumably along the southern extent of the $M_W~8.2$ rupture. The aftershocks are again sparse, but align as if they are occurring along strands of the fracture zones identified near the great earthquake's epicenter (Subcluster B). The overall aftershock alignment is oriented roughly 20-30N, which agrees with some of the aftershock fault plane strikes and with the estimated strike for the $M_W~8.2$ event, 17N. One pair of events, an $M_W$ 4.9 (2012/08/11) and $M_W$ 5.6 (2012/08/21), located near (92E, 0N) align closely in an east-west direction consistent with one of the GCMT fault planes. Thirty strongly correlated waveforms for these two events are azimuthally well distributed, but the distance between the events is very small (the estimate is 2-3 km), and so an east-west azimuth is preferred, but uncertain. Regardless, these events are almost certainly on the same structure, whether it aligns east west or north south.  

\subsection{Cluster E}
Observations are sparse for Cluster E (Fig. S11 in the electronic supplement), which provided six surface-wave relocatable aftershocks. The general trend of epicentroid locations is 325N (azimuth), which again is difficult to reconcile with the orientation of the GCMT faulting geometry estimates. The two easternmost events do not link to many other events, but do link to a well connected event slightly north. The location of the southernmost event may not be well located because the correlations do not form a tight, consistent cosine pattern. The easternmost event shows a pattern that indicates it is east of the trend formed by the other events.

\subsection{Clusters F and G}
Clusters F and G (Fig. S12 in the electronic supplement) are located within the Ninety-East Ridge. Only one pair of events align closely along one of the GCMT strikes. However, the relocation of one of the two events is not well constrained due to lack of event links. The depths of these events range from about 12 to 30 km. We observe some systematic Love and Rayleigh wave difference in misfit for the deepest event, but the overall sinusoidal pattern is robust. Whether the effect is a result of differing fault strikes or depth is unclear. In any case, the epicentroids form a T-shaped pattern with trends that align in a roughly north-south and ESE-WNW directions. The alignments do not correlate with the GCMT focal mechanism orientations, which remain consistent with the mechanisms throughout the aftershock area. 

\subsection{Cluster H}
Cluster H (Fig. S13 in the electronic supplement) is located north of clusters F and G, and consists of four relocated strike-slip events. Three of the four events have depth estimates (the fourth was fixed) and all are roughly at about 25~km depth. Again, the centroid locations do not form a simple consistent pattern with the fault orientations contained in the GCMT catalog. Even three events within 10-20~km of each other do not show any simple pattern. 

\section{Discussion}

The 2012 sequence aftershocks illuminate a complex fault system that includes several long linear epicentral patterns (Fig.~\ref{fig_usgsAShockMaps}) that may be associated with primary structures that failed during the great events of 2012. But when examined closely with information on individual event faulting geometry and precise relative locations, the moderate-magnitude earthquakes show that the trends represent structures more complex than simple through-going faults. The relationships between location patterns and GCMT faulting geometry (which we believe is accurate) is much more complex than what has been seen in typical oceanic transform environments, where the epicentroid location and GCMT faulting geometry strike estimates \citep{Cleveland:2013dm, Cleveland:2015gh} are quite consistent. 

Very conservatively, the great earthquakes in the 2012 sequence must have ruptured structures at least a few 10's of kilometers long, and probably those structures likely approached lengths of one hundred kilometers. But no such structures are well illuminated by moderate-size aftershocks. In the epicentral region of the mainshock, the alignments are inconsistent with estimated mainshock faulting geometry, which is of course complicated because it is an average of the moment tensors representing multiple sub-events involved in the earthquake \citep[e.g.][]{Duputel:2012ey, Yue:2012ks}. The aftershock faulting patterns are generally consistent with the $M_W$ 8.6 faulting geometry estimate, they do not align along expected directions. Nor are the aftershock trends aligned with the presumed trend of the fracture zones within the oceanic lithosphere in the region. We suspect that the aftershocks occurred along nearby lithospheric structures, generally correlated with mainshock strain release, but not along the main structures that failed during the $M_W$ 8.6. If the mainshock's fault or fault segments host few moderate-magnitude aftershocks, that might indicate relatively uniform slip on the mainshock rupture surface(s) (few stress concentrations requiring aftershock adjustment).

Near the $M_W$ 8.2 event, the situation may be more clear. In the USGS catalog, a 220 km long linear trend of events can be identified striking roughly 16N, compatible with the GCMT $M_W$ 8.2 event strike, and not too far from the strike of fracture zones in the region (Fig.~\ref{fig_clusterMap}). Our precise relocations suggest that the event strike and fracture-zone trends differ enough to require en echelon rupture of at least two fracture zone segments \citep[e.g.][]{Singh:2017gi}. The relocated epicentroids in Clusters B and D suggest that if the $M_W$ 8.2 occurred along these structures, perhaps as many as three strands of the fracture zone system participated, which is not a simple model of an earthquake rupture. Aftershocks along this trend suggest a rupture propagating 140~km to the SSW and possibly 80~km to the NNE. The southernmost seismicity in this pattern occurred later in the sequence (so it may simply represent aftershock area expansion \citep[e.g.][]{TAJIMA:1985ek}. The distance from the $M_W~8.2$ earthquake and the southernmost aftershock that we can locate is about 140~km. If the rupture propagated south for that distance, using the GCMT moment and assuming a mean shear modulus of 40~GPa would indicate a mean slip of 15 to 20 meters corresponding to a assumed seismogenic zone depth of 35 to 25 km, respectively. The P-wave back projections suggested a compact, but northward directed rupture for this event \citep{Yue:2012ks}. If we include the 80 km NNE of the $M_W$ 8.2 epicenter, then the rupture length is roughly 220~km and the slip estimates would decrease to 10 to 15 m. Although these slip estimates are reasonable, the overall rupture model is not simple and favors failure along this trend as opposed to a less obvious NNW rupture. In fact, \citet{Hill:2015iw} used seismic and geodetic observations to suggest that the slip along most active segments in the 2012 sequence, including the $M_W$ 8.2 event, occurred along the NNW striking structures conjugate to the fracture zones. Thus the aftershocks we analyze may simply illuminate segments of fractures zones stressed by the $M_W$ 8.2 slip.

The remaining epicentroid clusters show a similar pattern to the $M_W$~8.6. Some trends are apparent, but the alignment of the GCMT faulting geometry (candidate fault strikes) and the epicentroid trends is uncommon. Much of the aftershock sequence thus appears to be occurring along structures that likely have an en echelon pattern. Interestingly, the modified crust of the Ninety East ridge shows a similar complexity to the Wharton Basin lithosphere in that no large simple structures appear to dominate the deformation. Our locations suggest that the diffuse nature of the clustering aftershock activity is not an artifact of location, but is an intrinsic characteristic of the deformation during the 2012 earthquake sequence. Synthetic experiments indicate simple linear trends can be recovered with the surface-wave double-different technique if such trends exist.

\begin{figure}[!ht]
\centering
\includegraphics[width=\textwidth]{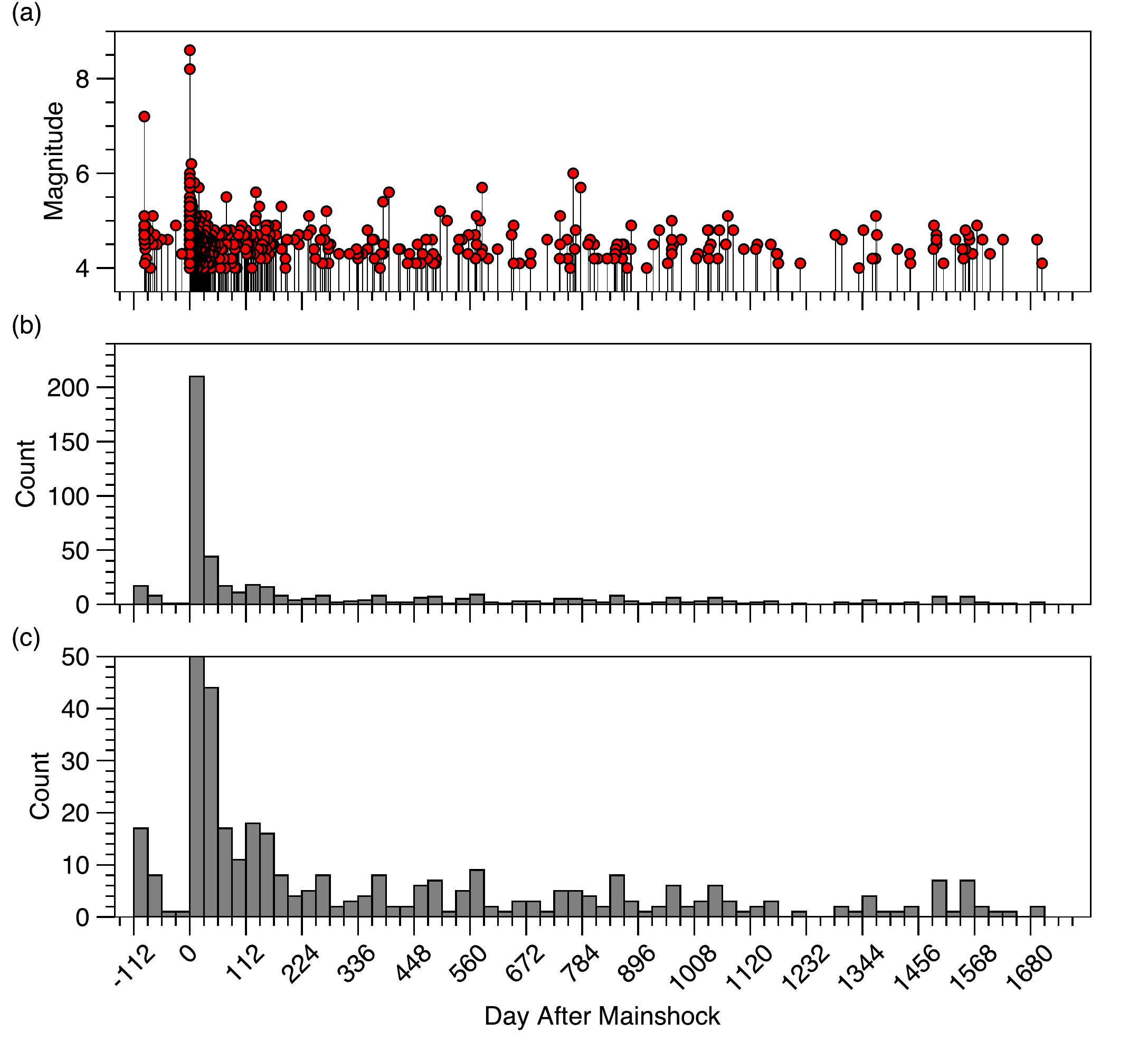}
\caption{Temporal variation of the seismicity from 2012 to 2017. Top panel shows a simple magnitude timeline starting 01 January, 2012. Time reference is the mainshock origin (12 April, 2012, 08:38). Bottom panel shows histogram of aftershock counts using bin widths of 28 days. The decay of activity was initially rapid (the mainshock bin is clipped), started out with some periodic behavior, and decayed slowly, similar to other intraplate earthquakes.}
\label{fig_usgsOmoriPlots}
\end{figure}

The magnitude and temporal patterns that have evolved over the last five years also show some interesting character. The temporal history of aftershock activity is summarized in Fig.~\ref{fig_usgsOmoriPlots}, which includes a magnitude time line and an Omori decay histogram. We used a bin width of 28 days for the event counts and the Omori display is clipped - the number of events in the first bin (including the great events) includes 400 events. Roughly 45\% of the aftershocks occurred in the first week, 58\% in the first month, and 78\% in the first year. The aftershock decay pattern is not steady, occasional moderate and strong events within the aftershock area produce aftershock sequences within the tail of the overall distribution. An interesting apparent periodicity seems to dominate the first few years of the sequence, which may suggest tidal aftershock triggering, but could also be an artifact. The decay of aftershocks for the 2012 sequences seems relatively sluggish. Following a rapid decay in the number of aftershocks in the first week to month, the decay in subsequent years is hard to detect (Fig.~\ref{fig_usgsOmoriPlots}). A slow decay is expected for intraplate earthquake since it is consistent with rate-state friction models \citep{DIETERICH:1994ku} or viscous dissipation processes \citep[e.g.][]{Ziv:2006dh} that suggest that the decay rate of aftershocks is inversely related to stress or tectonic loading processes. \citet{Stein:2009fa} used these ideas to explore the possibility that large intraplate earthquake aftershock sequences may in fact last for hundreds of years. 

To better quantify the decay of the 2012 earthquake aftershock sequence, we examined the temporal distribution of the aftershocks by fitting the modified Omori relationship \citep[reviewed in Appendix A][]{UTSU:1971us}. In Fig.~\ref{fig_OmorisLawLogLog}, we show the fit to the modified Omori Law for the 2012 $M_W$ 8.6 and for comparison, the 2005 $M_W$ 8.6 event subduction event (location indicated in Fig.~\ref{fig_gcmtMechanismBaseMap}). We sample the days (after the main shock) with a constant interval on a logarithmic scale. The USGS catalog was used for the modified Omori analysis. Based on the Gutenberg-Richter analysis (discussed in the following paragraph), we excluded events with magnitude less than 4. The three parameters $K$, $c$, and $p$ in the modified Omori relationship are estimated by minimizing the misfit between observations and predictions. The fit associated with the optimal parameters are satisfactory for both events as shown in Fig.~\ref{fig_OmorisLawLogLog}. The decay parameter $p$ of 1.0 for the 2012 event and of 0.8 for the 2005 event is typical comparing to other significant earthquakes \citep{Shcherbakov:2013ff}. The estimated parameters of the modified Omori Law are summarized in Table~\ref{tab_parameters}.

\begin{figure}[!ht]
\centering
\includegraphics[width=\textwidth]{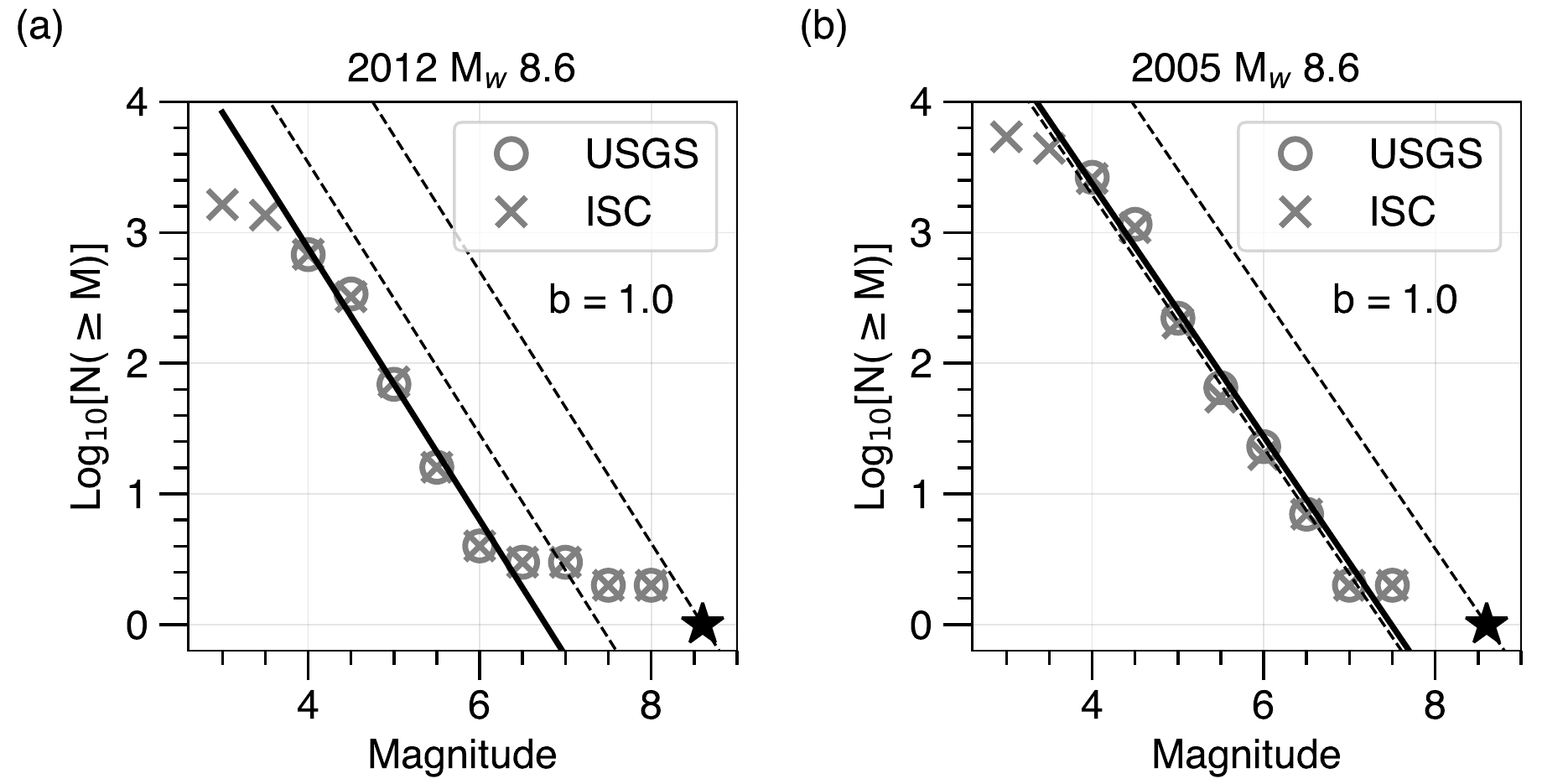}
\caption{Gutenberg-Richter plots for the 2012 $M_W$ 8.6 (a) and the 2005 $M_W$ 8.6 events (b). The solid line shows a linear fit to the observed frequency of earthquakes between magnitude 4.0 to 6.5 for (a) and between magnitude 4.0 to 7.5 for (b). The dashed lines project trends with slopes of -1 from the main shock magnitude, $M_W$ 8.6, and from 1.2 magnitude unit lower (accounting for a simple form of B\r{a}th's Law). Similar to most earthquakes in oceanic lithospheric the sequence is depleted in events as large as magnitude 6.0 for the 2012 earthquake sequence.}
\label{fig_usgsGRPlot}
\end{figure}

Fig.~\ref{fig_usgsGRPlot} includes a plot of the USGS and ISC magnitude distributions of the 2012 $M_W$ 8.6 sequence. Results for the 2005 $M_W$ 8.6 sequence are also included for comparison. We fit the cumulative magnitude distribution with a straight line as suggested by the Gutenberg-Richter relationship \citep[reviewed in Appendix A]{Gutenberg:1954iy}. We used a magnitude bin width of 0.5 units in the analysis (the results are similar for a bin width of 1.0). For the 2012 sequence, the magnitude distribution is not fit with a linear decay and a simple B\r{a}th's Law as is often the case for large continental intraplate earthquakes. Given the difference in the strain release pattern with a standard mainshock-aftershock sequence, it is not surprising that the usual pattern is not found. Between magnitudes of 4.0 to 6.5, the b-value is reasonably well approximated by unity (solid line in the figure). These numbers, however, are consistent with what would typically be observed following a magnitude 7 mainshock. The number of events with smaller magnitudes is substantially lower than that expected for the large events in the sequence. The dashed lines project from the mainshock magnitude and 1.2 unit less \citep[roughly accommodating B\r{a}th's Law,][]{Bath:1965dj} with a slope of minus one. Both lines suggest that we might have expected substantially more aftershocks than were observed. The pattern shows that similar to other earthquake sequences in oceanic lithosphere such as oceanic transforms  \citep[e.g.][]{Boettcher:2004ck} and deep earthquakes \citep{frohlich2006deep}, the 2012 aftershock sequence produced fewer aftershocks than typical. Only the large size of the two great earthquakes induced the hundreds of events observed across this broad region. The estimated parameters of the Gutenberg-Richter relationship are summarized in Table~\ref{tab_parameters}. For both the 2005 and 2012 sequences, the frequency-magnitude curve starts to deviate from the linear relationship for events with magnitude less than four, which indicates the catalogs are complete for magnitude larger or equal to four.

The 2012 great earthquakes were unusual and so it is no surprise that the aftershock patterns are also unusual. Perhaps the lack of aftershocks illuminating large structures is a result of a relatively uniform lithospheric structure (compared to plate boundary and continental regions) around what are likely relatively low-cumulative-offset faults \citep[e.g.][]{Hill:2015iw}. A lack of heterogeneity may result in a lack of focusing of stress onto smaller structures that commonly in other systems host aftershocks. Although not all, some have argued for super-shear rupture during the two great events, which is also consistent with observations of a lack of aftershocks in regions suspected of super-shear rupture along continental transform faults \citep[e.g.][]{Bouchon:2008jx}.

\begin{figure}[!ht]
\centering
\includegraphics[width=\textwidth]{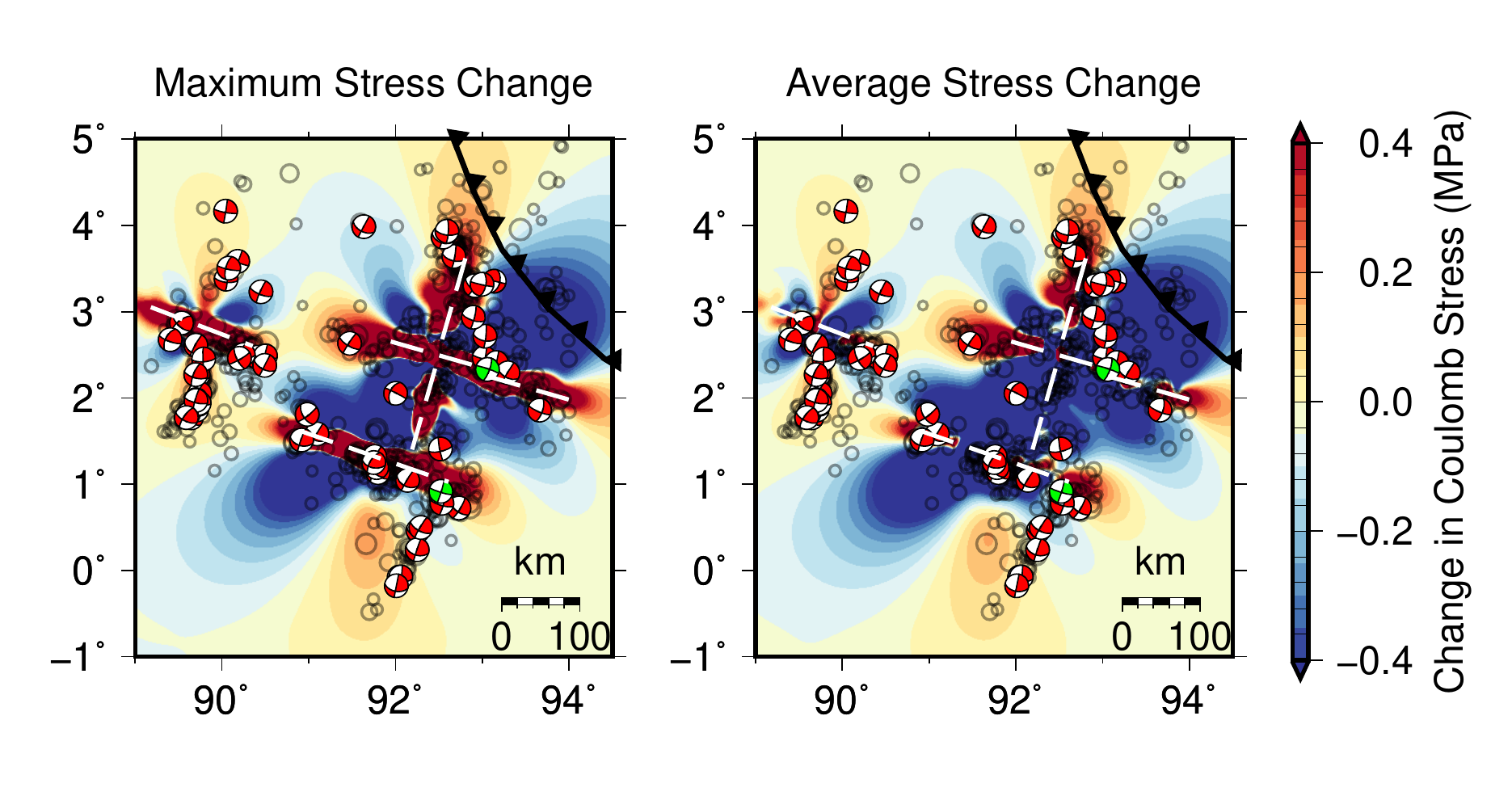}
\caption{Maximum (left) and average (right) static coulomb stress changes in the depth range of 0-30 km using the finite fault model from \citet{Yue:2012ks}. Circles identify USGS aftershock locations, the focal mechanisms identify GCMT faulting estimates shown at the centroid locations estimated in this study. Warm colors indicate regions of increased stress in a favorable orientation to produce faulting, cooler colors show regions where aftershock faulting is less likely.}
\label{fig_coulomb_stress}
\end{figure}

Finally, although complex, the distribution of aftershocks is compatible with the stress changes expected from models of the large-event slip distributions. First-order compatibility is expected since the aftershocks were used to identify likely rupture surfaces in the models. But consistency extends to regions off the fault segments as well. Figure~\ref{fig_coulomb_stress} is a plot of the maximum and average coulomb stress changes induced by the rupture model \citep[e.g.][]{Yue:2012ks}. The Coulomb stress changes were calculated using the Coulomb 3.3 package \citep{Lin:2004kf, Toda:2005bm} from USGS. We experimented with receiver faults corresponding to both stress-optimal orientations and to a specific orientation. When optimal orientations were chosen, the Coulomb stress changes do not agree well with the aftershock locations. Since the faulting geometry of most of these aftershocks is similar, we computed stress changes for a fault with a strike similar to the aftershock GCMT faulting geometry estimates. A range of friction parameters, dip angles, and the auxiliary planes were explored. For reasonable variations in these parameters, Coulomb stress changes show similar patterns. We present both the maximum and average stress changes (for a fault with a strike of 107, a dip of 75, and a rake of 180). We do not know a precise depth for many of the aftershocks, which could have occurred at different depths. Most aftershock depth estimates are within the upper 30~km. Both stress-change patterns are quite similar except very close to the mainshock rupture surfaces. The calculations show a reasonable agreement between predicted Coulomb stress increase and the moderate-magnitude aftershock locations. Many of the aftershocks that are too small for our location procedure also match with the Coulomb stress increase. Only a few exceptions exist, but consistent with our discussions above, not all regions expected to have increased stress produced aftershocks. The event produced fewer than expected aftershocks and very few align along the large-event ruptures.

\section{Conclusion}
The relative relocation of roughly 60 moderate-earthquake epicentroids suggests that aftershock faulting involved in the 2012 off-shore Indonesia sequence occurred in a region populated with many short fault segments. Unlike observed patterns along relatively mature transform faults, the relocated events seldom align along simple linear trends that are compatible with the strikes of the faults as estimated by the GCMT catalog. The lack of long, coherent structures suggests that the aftershocks are accommodating strain adjustments nearby, but off the larger structures that participated in the 2012 great earthquakes. The 2012 sequence has a relatively low number of aftershocks, which is similar to other fault ruptures in oceanic lithosphere. The decay of the aftershock rates are typical compared to other large events. Taken together, models of the rupture and the patterns in the aftershocks suggest that the eventual localization of deformation between the Indian and Australian plate is a complicated process still straining a large area beneath the northeastern Indian ocean.

\section{Acknowledgements}
This work was supported by the Defense Threat Reduction Agency under Award HDTRA1-11-1-0027. This material is based upon work partially supported by the U.S. Department of Energy, Office of Science, under contract number DE-AC05-00OR22725. The facilities of IRIS Data Services, and specifically the IRIS Data Management Center, were used for access the waveforms, related metadata, and/or derived products used in this study. IRIS Data Services are funded through the Seismological Facilities for the Advancement of Geoscience and EarthScope (SAGE) Proposal of the National Science Foundation under Cooperative Agreement EAR-1261681. The authors thank Monica Maceira and Philip Bingham for helpful comments. We acknowledge developers of Generic Mapping Tools \citep{Wessel:2013cb}, Obspy \citep{Beyreuther:2010gd, Megies:2011eb, Krischer:2015jm}, Numpy \citep{vanderWalt:2011jv}, and Matplotlib \citep{Hunter:2007ih} for sharing their packages. We also thank University of California San Diego for sharing the STRM15 topography data that were used as the background for several figures. This manuscript has been authored by UT-Battelle, LLC, under contract DE-AC05-00OR22725 with the US Department of Energy (DOE). The US government retains and the publisher, by accepting the article for publication, acknowledges that the US government retains a nonexclusive, paid-up, irrevocable, worldwide license to publish or reproduce the published form of this manuscript, or allow others to do so, for US government purposes. DOE will provide public access to these results of federally sponsored research in accordance with the DOE Public Access Plan (http://energy.gov/downloads/doe-public-access-plan). We thank an anonymous review and the Editor Rob Govers for constructive comments and suggestions. 

\pagebreak

\begin{figure}[!ht]
\centering
\includegraphics[width=\textwidth]{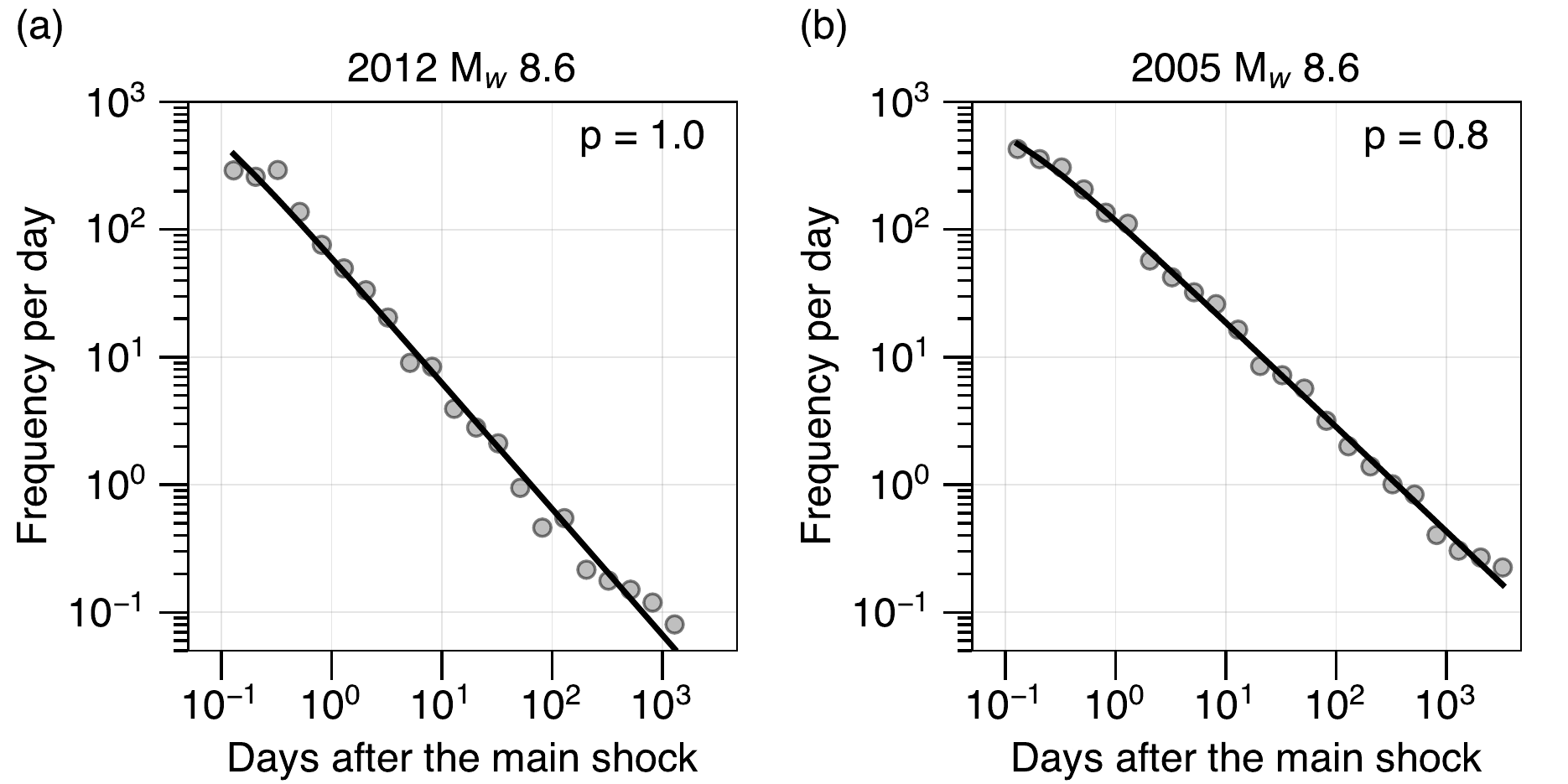}
\caption{Analysis results of the modified Omori's law for the 2012 M$_w$ 8.6 (a) and the 2005 M$_w$ 8.6 events (b). The gray dots show number of aftershocks per day as a function of days after the main shock. The black lines represent the best fit to the gray dots. }
\label{fig_OmorisLawLogLog}
\end{figure}


\clearpage

\begin{table}
\caption{Summary of the Estimated Parameters of the Empirical Relationships}
\label{tab_parameters}
\begin{tabular}{@{}ccccccc}
\hline
Event Date & Magnitude & a-value & b-value & $K$ & $C$ & $p$ \\
\hline
2005/03/28          & 8.6          & 7.2           & 1.0     & 124  & 0.07  & 0.8  \\
2012/04/11          & 8.6          & 7.0           & 1.0     & 61    & 0.02  & 1.0  \\
\hline
\end{tabular}
\end{table}

\clearpage



\appendix
\section{Gutenberg-Richter Relationship \& Modified Omori Law}

The frequency-magnitude statistics of  an earthquake sequence have been found to fit the Gutenberg-Richter relationship \citep{Gutenberg:1954iy}

\begin{equation}
	log_{10} N (\geq M) = a - b M
\end{equation}

where $N (\geq M)$ is the cumulative number of earthquakes with magnitudes larger than or equal to $M$. a and b are constants.

The temporal change of an earthquake sequence can be modeled by the modified Omori law \citep{UTSU:1971us}

\begin{equation}
	r(t) = \frac{K}{(t+C)^p}
\end{equation} 

where $r(t)$ is the occurrence rate of aftershocks at time $t$. $K$, $c$, and $p$ are constants.








\end{document}